 \documentclass[final,5p,times,twocolumn]{elsarticle}

\usepackage[nodots,nocompress]{numcompress}

\biboptions{super}

\journal{Solid State Communications}

\begin{document}

\begin{frontmatter}

%% Title, authors and addresses

%% use the tnoteref command within \title for footnotes;
%% use the tnotetext command for the associated footnote;
%% use the fnref command within \author or \address for footnotes;
%% use the fntext command for the associated footnote;
%% use the corref command within \author for corresponding author footnotes;
%% use the cortext command for the associated footnote;
%% use the ead command for the email address,
%% and the form \ead[url] for the home page:
%%
\title{Strain-controlled nonvolatile magnetization switching}

\author[WMI]{S.~Gepr\"{a}gs\corref{cor1}}
\ead{stephan.gepraegs@wmi.badw.de}

\author[WMI]{A.~Brandlmaier}

\author[WSI]{M.S.~Brandt}

\author[WMI,TUM]{R.~Gross}

\author[WMI]{S.T.B.~Goennenwein}

\cortext[cor1]{Corresponding author}

\address[WMI]{Walther-Mei{\ss}ner-Institut, Bayerische Akademie der
Wissenschaften, 85748 Garching, Germany}

\address[WSI]{Walter Schottky Institut, Technische Universit\"{a}t M\"{u}nchen, Am Coulombwall 4, 85748 Garching, Germany}

\address[TUM]{Physik-Department, Technische Universit\"{a}t M\"{u}nchen, 85748 Garching, Germany}

\begin{abstract}
%% Text of abstract
We investigate different approaches towards a nonvolatile switching of the remanent magnetization in single-crystalline ferromagnets at room temperature via elastic strain using ferromagnetic thin film/piezoelectric actuator hybrids. The piezoelectric actuator induces a voltage-controllable strain along different crystalline directions of the ferromagnetic thin film, resulting in modifications of its magnetization by converse magnetoelastic effects. We quantify the magnetization changes in the hybrids via ferromagnetic resonance spectroscopy and superconducting quantum interference device magnetometry. These measurements demonstrate a significant strain-induced change of the magnetization, limited by an inefficient strain transfer and domain formation in the particular system studied. To overcome these obstacles, we address practicable engineering concepts and use a model to demonstrate that a strain-controlled, nonvolatile magnetization switching should be possible in appropriately engineered ferromagnetic/piezoelectric actuator hybrids.
\end{abstract}

\begin{keyword}
A. Multiferroic Hybrids \sep D. Magnetostriction \sep E. Ferromagnetic Resonance \sep E. SQUID magnetometry
%% keywords here, in the form: keyword \sep keyword

%% MSC codes here, in the form: \MSC code \sep code
%% or \MSC[2008] code \sep code (2000 is the default)

\end{keyword}

\end{frontmatter}

%%
%% Start line numbering here if you want
%%
% \linenumbers

%% main text
\section{Introduction}
\label{sec:Intro}

In magnetoelectric multiferroics, where the ferromagnetic and ferroelectric order parameters are coupled, an electric-field control of the magnetic properties becomes possible.\cite{Fiebig:JPhysD:38:2005,Eerenstein:Nature:442:2006,Ramesh:NatMater:6:2007} This opens the way for appealing novel magnetization control schemes in future spintronic devices.\cite{Zhao:NatMater:5:2006} Unfortunately, single-phase multiferroics with strong magnetoelectric coupling remain rare.\cite{Hill:JPhysChemB:104:2000,Eerenstein:Nature:442:2006} Attractive alternatives are composite material systems made from ferroelectric and ferromagnetic compounds.\cite{Gepraegs:PhilosMagLett:87:2007,Nan:JAP:103:2008,Ma:AdvMater:23:2011,Vaz:JPhysCondMatt:24:2012} In such systems, an electric-field control of magnetism can be realized using electric field effects in carrier-mediated ferromagnets,\cite{Stolichnov:NatMater:7:2008,Vaz:AdvMater:22:2010} or exchange coupling at ferromagnetic/multiferroic interfaces.\cite{Borisov:PRL:94:2005,Chu:NatMater:7:2008} A third, powerful approach relies on strain-mediated, indirect magnetoelectric coupling in ferromagnetic/ferroelectric hybrid systems. In recent years, these hybrids were mostly fabricated by depositing ferromagnetic thin films on ferroelectric substrates.\cite{Lee:APL:82:2003,Eerenstein:NatMater:6:2007,Sahoo:PRB:76:2007,Thiele:PRB:75:2007,Israel:APL:93:2008,Liu:AdvFunctMater:19:2009,Gepraegs:APL:96:2010,Liu:JAP:107:2010,Gepraegs:PRB:86:2012,Lahtinen:SciRep:2:2012,Streubel:PRB:87:2013,Ghidini:NatureCom:4:2013} Another approach to realize a strain-mediated control of the magnetization is to fabricate ferromagnetic thin film/piezoelectric actuator hybrids by either depositing or cementing ferromagnetic thin films onto commercially available $\mathrm{Pb}\left(\mathrm{Zr}_{x}\mathrm{Ti}_{1-x}\right)\mathrm{O}_{3}$ (PZT) multilayer piezoelectric actuator stacks [cf.~Fig.~\ref{fig:piezo-coordsyst}(a)].\cite{Brandlmaier:PRB:77:2008,Bihler:PRB:78:2008,Overby:APL:92:2008,Rushforth:PRB:78:2008,Weiler:NJP:11:2009,Brandlmaier:JAP:110:2011} In these hybrids, the application of a voltage to the piezoelectric actuator results in a deformation, which is transferred to the overlaying ferromagnetic thin film, changing its magnetic anisotropy due to the converse magnetoelastic effect.

In this paper, we report on two different experimental approaches towards a strain-mediated, nonvolatile, voltage-controlled magnetization switching in the complete absence of magnetic fields. They are based on ferromagnetic thin film/piezoelectric actuator hybrids using $\mathrm{Fe}_{3}\mathrm{O}_{4}$ as the ferromagnet. Our experiments show that a significant modification of the magnetic anisotropy is possible via voltage-controlled strain. This work extends our previous studies on ferromagnetic/ferroelectric hybrids,\cite{Brandlmaier:PRB:77:2008,Bihler:PRB:78:2008,Weiler:NJP:11:2009,Gepraegs:APL:96:2010} where we achieved a reversible reorientation of the magnetization by up to $90^{\circ}$ in Ni based hybrids. However, a true switching of the magnetization between two (or more) remanent states solely by means of an electric field induced strain has not been realized experimentally up to now.\cite{Pertsev:PRB:78:2008,Pertsev:APL:95:2009,Hu:PRB:80:2009,Hu:JAP:107:2010,Hu:JAP:109:2011,Iwasaki:JMMM:240:2002}  

\section{The spin-mechanics concept}
\label{sec:Concept}

\begin{figure}
\includegraphics[width=0.95\columnwidth]{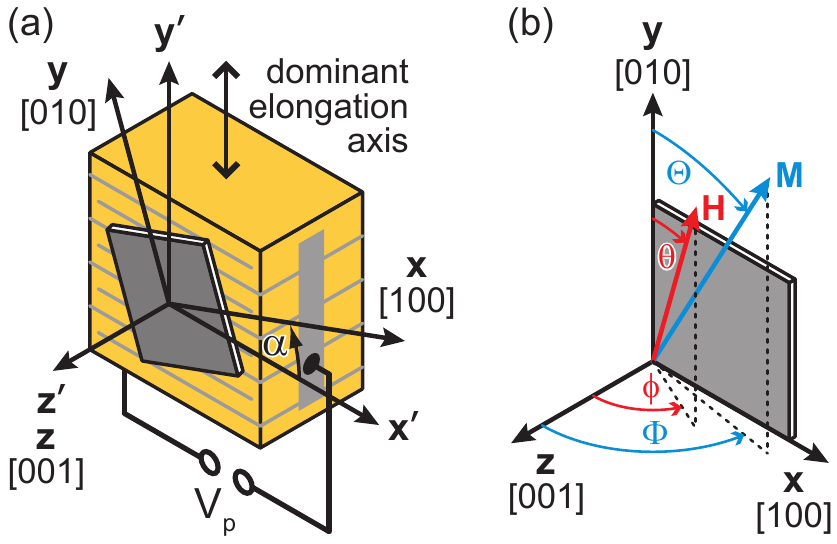}
\caption{(a) Schematic illustration of a $\mathrm{Fe}_{3}\mathrm{O}_{4}$ thin film/piezoelectric
actuator hybrid. The coordinate system of the thin film and the actuator enclosing an angle $\alpha$ are denoted by $\left\{ \mathbf{x},\mathbf{y},\mathbf{z}\right\}$ and $\left\{ \mathbf{x}^{\prime},\mathbf{y}^{\prime},\mathbf{z}^{\prime}\right\}$, respectively. (b) Orientation of the magnetic field $\mathbf{H}\left(H,\theta,\phi\right)$ and the magnetization $\mathbf{M}\left(M_{\mathrm{s}},\Theta,\Phi\right)$ with respect to the crystallographic axes $\left\langle 100\right\rangle $ of the $\mathrm{Fe}_{3}\mathrm{O}_{4}$ thin film.} 
\label{fig:piezo-coordsyst}
\end{figure}

The orientation of a well-defined homogeneous magnetization in a ferromagnet depends on external mechanical stress due to magnetostriction.\cite{Sander:RPP:62:1999,Chikazumi:book:1997} We exploit this so-called spin-mechanics scheme to control the magnetic anisotropy in $\mathrm{Fe}_{3}\mathrm{O}_{4}$ thin films cemented on $\mathrm{Pb}\left(\mathrm{Zr}_{x}\mathrm{Ti}_{1-x}\right)\mathrm{O}_{3}$ (PZT) multilayer piezoelectric actuator stacks [cf.~Fig.~\ref{fig:piezo-coordsyst}(a)]. In particular, we compare different hybrids fabricated by cementing $\mathrm{Fe}_{3}\mathrm{O}_{4}$ thin films with different angles $\alpha$ between the crystallographic axes $\left\{ \mathbf{x},\mathbf{y}\right\}$ of the film and the principal elongation axes $\left\{\mathbf{x}^{\prime},\mathbf{y}^{\prime}\right\}$  of the actuator with $\mathbf{z}\parallel\mathbf{z}^{\prime}$.

%%%%%%%%%%%%%%%%%%%%%%%%%%%%%%%%%%%%%%% Fig.1 %%%%%%%%%%%%%%%%%%%%%%%%%%%%%%%%%%%%%%%%%%%%%%
%%%%%%%%%%%%%%%%%%%%%%%%%%%%%%%%%%%%%%% Fig.1 %%%%%%%%%%%%%%%%%%%%%%%%%%%%%%%%%%%%%%%%%%%%%%

The application of a voltage $V_{\mathrm{p}}>0$ $\left(V_{\mathrm{p}}<0\right)$ to the piezoelectric actuator causes an elongation $\epsilon_{2}^{\prime}>0$ (contraction $\epsilon_{2}^{\prime}<0$) along the actuator's dominant elongation axis $\mathbf{y}^{\prime}$, which is due to elasticity accompanied by a contraction (elongation) along the two orthogonal directions $\mathbf{x}^{\prime}$ and $\mathbf{z}^{\prime}$ [cf.~Fig.~\ref{fig:piezo-coordsyst}(a)]. This leads to a change of the strain state $\mathrm{\epsilon}$ of the $\mathrm{Fe}_{3}\mathrm{O}_{4}$ thin film elastically clamped onto the piezoelectric actuator. This causes a modification of the magnetic anisotropy, and thus alters the direction of the magnetization $\mathbf{M}$. In a macrospin model, the magnetization $\mathbf{M}$ of the $\mathrm{Fe}_{3}\mathrm{O}_{4}$ thin film described by $\mathbf{M}\left(M_{\mathrm{s}},\Theta,\Phi\right)=M_{\mathrm{s}}\mathbf{m}\left(\Theta,\Phi\right)$ aligns in such a way that the free energy density $F$ takes its minimum value in equilibrium. Here, $m_{x}=\sin\Theta\sin\Phi$, $m_{y}=\cos\Theta$, and $m_{z}=\sin\Theta\cos\Phi$ [cf.~Fig.~\ref{fig:piezo-coordsyst}(b)] are directional cosines and $M_{\mathrm{s}}$ the saturation magnetization. The orientation of the magnetization $\mathbf{m}\left(\Theta,\Phi\right)$ can be calculated in the framework of a single domain model by using a phenomenological thermodynamic model based on the free energy density
\begin{equation}
F=F_{\mathrm{Zeeman}}+F_{\mathrm{u,eff}}^{001}+F_{\mathrm{mc}}+F_{\mathrm{el}}+F_{\mathrm{me}}
\label{eq:F-tot_general}
\end{equation} 
with the Zeeman energy density $F_{\mathrm{Zeeman}}=-\mu_{0}M_{\mathrm{s}}\mathbf{m}\left(\Theta,\Phi\right)H\mathbf{h}\left(\theta,\phi\right)$, the effective uniaxial anisotropy contribution $F_{\mathrm{u,eff}}^{001}=\frac{1}{2}\mu_{0}M_{\mathrm{s}}^{2}m_{z}^{2}+K_{\mathrm{u}}^{001}m_{z}^{2}$, which comprises the demagnetization contribution and the uniaxial contribution $K_{\mathrm{u}}^{001}$ resulting from the pseudomorphic growth of the ferromagnetic thin film, the first-order magnetocrystalline anisotropy contribution $F_{\mathrm{mc}}=K_{\mathrm{c}}\left(m_{x}^{2}m_{y}^{2}+m_{y}^{2}m_{z}^{2}+m_{z}^{2}m_{x}^{2}\right)$ with the cubic anisotropy constant $K_{\mathrm{c}}$, the elastic energy density\cite{MatrixNotation}   $F_{\mathrm{el}}=\frac{1}{2}c_{11}\left(\epsilon_{1}^{2}+\epsilon_{2}^{2}+\epsilon_{3}^{2}\right)+c_{12}\left(\epsilon_{1}\epsilon_{2}+\epsilon_{2}\epsilon_{3}+\epsilon_{1}\epsilon_{3}\right)+\frac{1}{2}c_{44}\left(\epsilon_{4}^{2}+\epsilon_{5}^{2}+\epsilon_{6}^{2}\right)$, and the magnetoelastic contribution 
\begin{eqnarray}
F_{\mathrm{me}} & = & B_{1}\left[\epsilon_{1}\left(m_{x}^{2}-\frac{1}{3}\right)+\epsilon_{2}\left(m_{y}^{2}-\frac{1}{3}\right)+\epsilon_{3}\left(m_{z}^{2}-\frac{1}{3}\right)\right]\nonumber \\
 &  & +B_{2}\left(\epsilon_{4}m_{y}m_{z}+\epsilon_{5}m_{x}m_{z}+\epsilon_{6}m_{x}m_{y}\right)\,.
\label{eq:F-magel_general}
\end{eqnarray}
The magnetoelastic coupling coefficients $B_{1}$ and $B_{2}$ can be written as a function of the magnetostrictive constants $\lambda_{100}$ and $\lambda_{111}$, which yields $B_{1}=-\frac{3}{2}\lambda_{100}\left(c_{11}-c_{12}\right)$ and $B_{2}=-3\lambda_{111}c_{44}$. Here we use bulk values for the magnetostrictive constants ($\lambda_{100}=-19.5\times10^{-6}$ and $\lambda_{111}=+77.6\times10^{-6}$) as well as for the elastic stiffness constants $c_{ij}$ $(c_{11}=27.2\times10^{10}$\,N/m$^{2}$, $c_{12}=17.8\times10^{10}$\,N/m$^{2}$, and $c_{44}=6.1\times10^{10}$\,N/m$^{2})$.\cite{Gorter:ProcIRE:43:1955,Bickford:PR:99:1955,Schwenk:EPJB:13:2000}

To determine the modification of the magnetic anisotropy caused by strain effects induced by the piezoelectric actuator, we first derive the strain tensor $\mathrm{\epsilon}$ of the $\mathrm{Fe}_{3}\mathrm{O}_{4}$ thin film. In the $\left\{\mathbf{x}^{\prime},\mathbf{y}^{\prime},\mathbf{z}^{\prime}\right\}$ coordinate system, the strain components $\epsilon_{4}^{\prime}$, $\epsilon_{5}^{\prime}$, and $\epsilon_{6}^{\prime}$ vanish, since no shear strains are present. 
%%%%%%%%%%%%%%%%%%%%%%%%%%%%%%%%%%%%%%% Fig.2 %%%%%%%%%%%%%%%%%%%%%%%%%%%%%%%%%%%%%%%%%%%%%%
\begin{figure}[tbh]
\begin{centering}
\includegraphics[width=0.95\columnwidth]{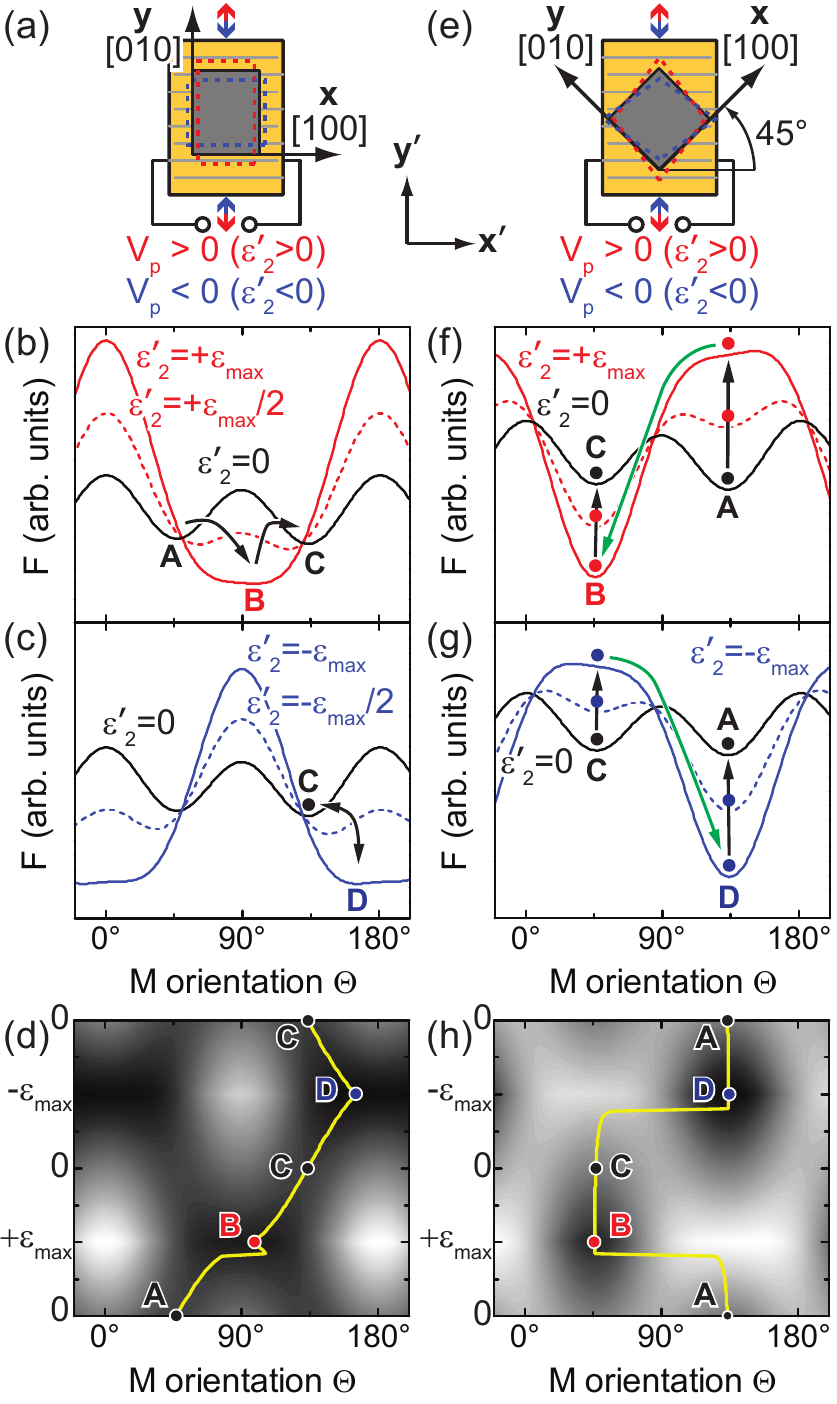}
\end{centering}
\caption{(a)--(d) Stress applied to a cubic thin film along the in-plane crystallographic $\left\langle 100\right\rangle $ axes ($\alpha=0{}^{\circ}$). (b)--(d) Corresponding free energy density contours $F\left(\Theta,\epsilon_{2}^{\prime}\right)$, with capital letters indicating the equilibrium magnetization orientations. The full yellow line in (d) traces the minimum of $F$. Forward switching occurs $\left(\mathrm{A} \rightarrow \mathrm{C}\right)$, while a back switching $\left(\mathrm{C} \rightarrow \mathrm{A}\right)$ is not possible. (e)--(h) Deformation of a cubic crystal along the in-plane $\left\langle 110\right\rangle $ axes ($\alpha=45{}^{\circ}$). Both a forward and a back switching is feasible. The green arrows illustrate the discontinuous change of the magnetization orientation.}
\label{fig:Illustration-FreeE}
\end{figure}
%%%%%%%%%%%%%%%%%%%%%%%%%%%%%%%%%%%%%%% Fig.2 %%%%%%%%%%%%%%%%%%%%%%%%%%%%%%%%%%%%%%%%%%%%%%
Furthermore, as the thin film is clamped to the piezoelectric actuator, the in-plane strains $\epsilon_{1}^{\prime}$ and $\epsilon_{2}^{\prime}$ are not independent. Due to the actuator's elastic properties, these strain components are related via the Poisson ratio $\nu=0.45$ according to $\epsilon_{1}^{\prime}=-\nu\epsilon_{2}^{\prime}$. To obtain the strain tensor $\mathbf{\epsilon}$ in the coordinate system of the $\mathrm{Fe}_{3}\mathrm{O}_{4}$ thin film $\left\{\mathbf{x},\mathbf{y},\mathbf{z}\right\}$, we apply a tensor transformation as described in detail in Refs.~\cite{Sander:RPP:62:1999,Nye:book:1985}. The strain components $\epsilon_{i}$ $\left(i=3,4,5\right)$ can then be deduced according to the mechanical equilibrium condition $\sigma_{i}=\partial F/\partial\epsilon_{i}=0$ $\left(i=3,4,5\right)$. With this relation, we finally obtain $\mathbf{\epsilon}$ as a function of $\epsilon_{2}^{\prime}$, neglecting comparably small magnetoelastic terms:
\begin{equation}
\mathbf{\epsilon}=\left(\begin{array}{c}
-\frac{1}{2}\left[-1+\nu+\left(1+\nu\right)\cos\left(2\alpha\right)\right]\epsilon_{2}^{\prime}\\
\frac{1}{2}\left[1-\nu+\left(1+\nu\right)\cos\left(2\alpha\right)\right]\epsilon_{2}^{\prime}\\
-\frac{c_{12}}{c_{11}}\left(1-\nu\right)\epsilon_{2}^{\prime}\\
0\\
0\\
\left(1+\nu\right)\sin\left(2\alpha\right)\epsilon_{2}^{\prime}
\end{array}\right)\,.
\label{eq:strain_general}
\end{equation}

Now we are in a position to derive the magnetization orientation $\mathbf{m}\left(\Theta,\Phi\right)$ by tracing the minimum of the total free energy density $F$ as a function of $\epsilon_{2}^{\prime}$, which can be controlled by $V_{\mathrm{p}}$. The corresponding evolution is calculated by minimizing Eq.~(\ref{eq:F-tot_general}) with respect to the orientation of the magnetization $\Theta$. Since the strain induced in the ferromagnetic thin film is of the order of $10^{-3}$ in our hybrid structures, the magnetoelastic energy contribution $F_{\mathrm{me}}$ will not overcome the demagnetization energy in $\mathrm{Fe}_{3}\mathrm{O}_{4}$ thin films. Thus, the magnetization remains in-plane in case of zero magnetic field, which results in $\Phi=90^{\circ}$. 

To illustrate the concept of a strain-induced, nonvolatile magnetization switching in zero magnetic and electric fields, Fig.~\ref{fig:Illustration-FreeE} exemplary shows free energy density $F(\Theta,\epsilon_{2}^{\prime})$ contours within the film plane for $\alpha=0{}^{\circ}$ [Fig.~\ref{fig:Illustration-FreeE}(b)--(d)] and $\alpha=45{}^{\circ}$ [Fig.~\ref{fig:Illustration-FreeE}(f)--(h)]. In both cases, the induced uniaxial strain is symmetric with respect to the crystallographic axes of the cubic ferromagnetic thin film. This results in two energetically equivalent minima in the free energy density $F$, which forces domain formation. To lift this degeneracy a small uniaxial magnetic anisotropy contribution in the film plane is introduced in the simulations given by $F_{\mathrm{u}}^{\mathrm{ip}}=K_{\mathrm{u}}^{\mathrm{ip}}\left(m_{x}\sin\Theta_{\mathrm{u}}+m_{y}\cos\Theta_{\mathrm{u}}\right)^{2}$ with the uniaxial anisotropy constant $K_{\mathrm{u}}^{\mathrm{ip}}$. For illustration purposes, we here use $\Theta_{\mathrm{u}}=10{}^{\circ}$ and $K_{\mathrm{u}}^{\mathrm{ip}}>0$ with $\left|K_{\mathrm{u}}^{\mathrm{ip}}/K_{\mathrm{c}}\right|=1/15$. To meet the experimental conditions of $\mathrm{Fe}_{3}\mathrm{O}_{4}$ thin films, we choose $K_{\mathrm{u}}^{001}>0$ and $K_{\mathrm{c}}<0$.

In case of $\alpha=0^{\circ}$, the ferromagnetic thin film is elongated and contracted along the cubic axes ($\mathbf{x}^{\prime}||\mathbf{x}$ and $\mathbf{y}^{\prime}||\mathbf{y}$) [cf.~Fig.~\ref{fig:Illustration-FreeE}(a)] and thus no shear strains appear ($\epsilon_{6}=0$) [cf.~Eq.~(\ref{eq:strain_general})]. Starting at $\epsilon_{2}^{\prime}=0$ ($V_{\mathrm{p}}=0$\,V) [cf.~black line in Fig.~\ref{fig:Illustration-FreeE}(b)], the magnetization orientation $\Theta$ is aligned along a magnetically easy axis,~e.g.,~at $\Theta=47^{\circ}$ (point A). Upon increasing $\epsilon_{2}^{\prime}$ ($V_{\mathrm{p}}>0$\,V), the magnetically easy axis and thus the magnetization orientation $\Theta$ continuously rotates towards $\Theta=98^{\circ}$ (point B). The corresponding free energy density contour for $\epsilon_{2}^{\prime}=+\epsilon_{\mathrm{max}}$ [cf.~red line in Fig.~\ref{fig:Illustration-FreeE}(b)] is calculated assuming $B_{1}\epsilon_{\mathrm{max}}/K_{\mathrm{c}}=3/5$, which corresponds to $\epsilon_{\mathrm{max}}=1\times10^{-3}$ in case of $\mathrm{Fe}_{3}\mathrm{O}_{4}$ thin films. By decreasing $\epsilon_{2}^{\prime}$ back to $0$, the magnetization orientation continuously rotates to the energetically stable direction $\Theta=133^{\circ}$ (point C) at $\epsilon_{2}^{\prime}=0$. This demonstrates that a reorientation of the magnetization by about $86^{\circ}$ is feasible. To check the possibility to reorient the magnetization orientation to the initial configuration (point A), $\epsilon_{2}^{\prime}$ is inverted. Figure~\ref{fig:Illustration-FreeE}(c) discloses that the easy axis gradually rotates from $\Theta=133{}^{\circ}$ (point C) to $\Theta=165{}^{\circ}$ (point D) by inducing $\epsilon_{2}^{\prime}=-\epsilon_{\mathrm{max}}$. However, upon reducing $\epsilon_{2}^{\prime}$ back to $0$, the easy axis rotates back to $\Theta=133{}^{\circ}$ (point C). Thus, the magnetization remains in point C and a further strain-induced switching process is not possible. Consequently, the configuration $\alpha=0{}^{\circ}$ allows for a single, irreversible, and nonvolatile magnetization switching. The whole reorientation process of the magnetization for $\alpha=0{}^{\circ}$ is shown in Fig.~\ref{fig:Illustration-FreeE}(d), which displays the free energy density surface $F(\Theta,\epsilon_{2}^{\prime})$. The full yellow line traces the minimum of $F$. By applying the sequence $0 \rightarrow +\epsilon_{\mathrm{max}} \rightarrow 0 \rightarrow -\epsilon_{\mathrm{max}} \rightarrow 0$ for $\epsilon_{2}^{\prime}$, the remanent magnetization aligns along $\mathrm{A} \rightarrow \mathrm{B} \rightarrow \mathrm{C} \rightarrow \mathrm{D} \rightarrow \mathrm{C}$.

In contrast to $\alpha=0^{\circ}$, the configuration with $\alpha=45^{\circ}$ leads to a finite shear strain component $\epsilon_{6}\neq0$, since the piezoelectric actuator exerts stress along the in-plane $\left\langle 110\right\rangle $ directions of the ferromagnetic thin film [cf.~Fig.~\ref{fig:Illustration-FreeE}(e)]. For simplicity, we assume $\left|B_{1}\epsilon_{\mathrm{max}}/K_{\mathrm{c}}\right|=\left|B_{2}\epsilon_{\mathrm{max}}/K_{\mathrm{c}}\right|$. At the beginning ($\epsilon_{2}^{\prime}=0$) [cf. black line in \ref{fig:Illustration-FreeE}(f)], the magnetization orientation $\Theta$ is aligned along $133^{\circ}$ (point A). While increasing $\epsilon_{2}^{\prime}$, the easy axis basically retains its initial orientation. However, the free energy density minimum gradually transforms into a maximum. Upon a certain critical induced strain $\epsilon^{\mathrm{sw}}$, the easy axis changes discontinuously to $\Theta=46^{\circ}$ (point B), indicating an abrupt magnetization switching [cf. green arrow in \ref{fig:Illustration-FreeE}(f)]. The orientation of the easy axis essentially stays along $\Theta=46{}^{\circ}$ while reducing $\epsilon_{2}^{\prime}$ back to $0$ (point C). Subsequently, we continuously increase the inverted induced strain $\epsilon_{2}^{\prime}<0$ [Fig.~\ref{fig:Illustration-FreeE}(g)]. Starting from point C the easy axis abruptly rotates to $\Theta=133^{\circ}$ (point D). This magnetization orientation remains unchanged, while increasing $\epsilon_{2}^{\prime}$ back to zero again. Thus, in case of $\alpha=45^{\circ}$, a reorientation of the magnetization back to the initial state is possible, which demonstrates that a reversible, nonvolatile magnetization switching in the absence of a magnetic field is possible. The switching of the magnetization from point A to point C and back to point A upon applying the strain sequence $0 \rightarrow + \epsilon_{\mathrm{max}} \rightarrow 0 \rightarrow - \epsilon_{\mathrm{max}} \rightarrow 0$ is further illustrated in Fig.~\ref{fig:Illustration-FreeE}(h).

\section{Towards a nonvolatile magnetization switching via strain in experiment}
\label{sec:Exp-FMR}

As described in the previous section, a nonvolatile magnetization switching is theoretically possible in $\mathrm{Fe}_{3}\mathrm{O}_{4}$ thin film/piezoelectric actuator hybrid structures. In the following, we discuss two hybrids corresponding to the configurations discussed in Section~\ref{sec:Concept}. The hybrids are based on the same (001)-oriented, 44\,nm thick $\mathrm{Fe}_{3}\mathrm{O}_{4}$ film grown on a MgO (001) substrate by laser-MBE. After the deposition the thin film sample was cut into two pieces, which were cemented onto the piezoelectric actuators in such a way that stress is either exerted along the $\left\langle 100\right\rangle $ crystal axes ($\alpha=0^{\circ}$) or along $\left\langle 110\right\rangle $ ($\alpha=45^{\circ}$). The fabrication process of the thin film/piezoelectric actuator hybrid structure is described in detail in Ref.~\cite{Brandlmaier:PRB:77:2008}. The samples thus obtained are referred to as hybrid~$\left\langle 100\right\rangle $ and hybrid~$\left\langle 110\right\rangle $, respectively. The magnetic anisotropy of the $\mathrm{Fe}_{3}\mathrm{O}_{4}$ thin film was determined by angular-dependent ferromagnetic resonance (FMR) spectroscopy at constant actuator voltages $V_{\mathrm{p}}$ with the magnetic field applied in the film plane $\mathbf{h}\left(\theta,\phi=90^{\circ}\right)$ at room temperature.\cite{Brandlmaier:PRB:77:2008} 

\begin{figure}
\begin{centering}
\includegraphics[width=0.95\columnwidth]{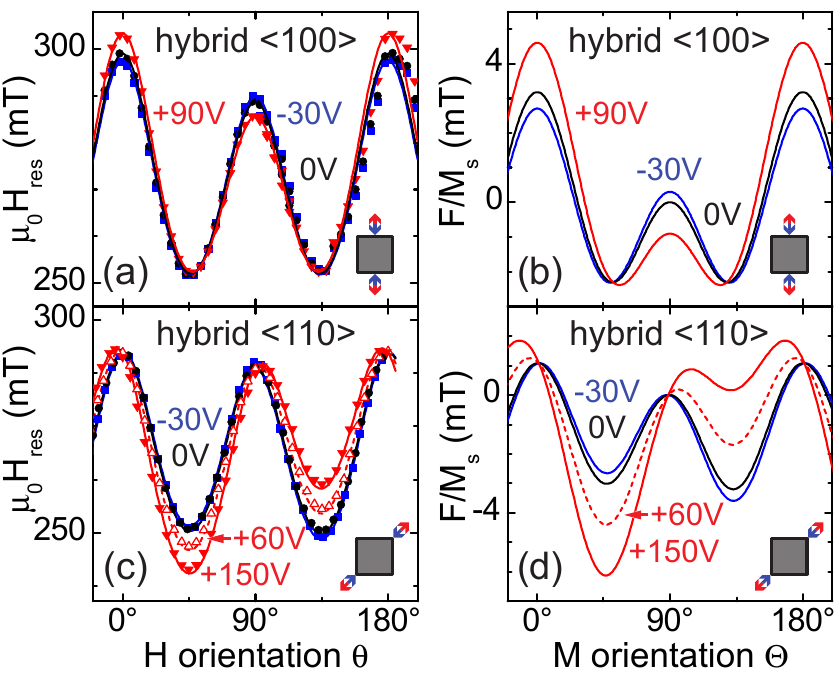}
\end{centering}
\caption{FMR fields $\mu_{0}H_{\mathrm{res}}$ for a rotation of the magnetic field in the film plane $\mathbf{h}\left(\theta,\phi=90^{\circ}\right)$ as a function of
$V_{\mathrm{p}}$ (symbols) for the hybrid~$\left\langle 100\right\rangle $ (a) and the hybrid~$\left\langle 110\right\rangle $ (c). The lines depict the simulated FMR fields. (b),~(d) Calculated free energy density contours as a function of the magnetization orientation $\Theta$ in the film plane at zero external magnetic field.}
\label{fig:FMR-FreeE}
\end{figure}

For the hybrid~$\left\langle 100\right\rangle $ ($\alpha=0^{\circ}$), the evolution of the obtained FMR fields $\mu_{0}H_{\mathrm{res}}\left(\theta\right)$ as a function of the external magnetic field orientation $\theta$ reveals a superposition of a cubic magnetic anisotropy with a uniaxial one [cf.~Fig.~\ref{fig:FMR-FreeE}(a)].\cite{Brandlmaier:PRB:77:2008} For a quantitative simulation of the experimental data, the FMR angular dependence is simulated according to Eq.~(\ref{eq:F-tot_general}).\cite{Farle:RPP:61:1998,Brandlmaier:PRB:77:2008}. The best agreement between the FMR fields for $V_{\mathrm{p}} = 0$\,V observed in experiment [cf.~black symbols in Fig.~\ref{fig:FMR-FreeE}(a)] and simulation [cf.~black line in Fig.~\ref{fig:FMR-FreeE}(a)] was obtained by using the voltage-independent anisotropy fields $K_{\mathrm{u,eff}}^{001}/M_{\mathrm{s}}=80.2$\,mT, $K_{\mathrm{c}}/M_{\mathrm{s}}=-14.9$\,mT, and $K_{\mathrm{u}}^{\mathrm{ip}}/M_{\mathrm{s}}=3.2$\,mT. An additional uniaxial contribution in the film plane $F_{\mathrm{u}}^{\mathrm{ip}}$ with $\theta_{\mathrm{u}}=0{}^{\circ}$, which is not observed in the as-grown $\mathrm{Fe}_{3}\mathrm{O}_{4}$ thin film and caused by an anisotropic thermal expansion during the curing process, has to be included in the free energy density $F$. For $V_{\mathrm{p}}\neq 0$\,V,~i.e.,~$\epsilon_{2}^{\prime}\neq0$, the FMR fields $\mu_{0}H_{\mathrm{res}}$ are modeled by using $\epsilon_{2}^{\prime}$ as fit parameter [cf.~red and blue lines in Fig.~\ref{fig:FMR-FreeE}(a)]. The derived strain $\Delta\epsilon_{2}^{\prime}=\epsilon_{2}^{\prime}(+90$\,V$)-\epsilon_{2}^{\prime}(-30$\,V$)=0.23\times 10^{-3}$ induced in the ferromagnetic thin film amounts to only about 27\% of the nominal stroke of $\Delta\epsilon_{2}^{\mathrm{ideal}}=0.87\times 10^{-3}$ of the piezoelectric actuator.\cite{piezomechanik:booklet:LowVoltStacks:2010} This is most likely caused by an imperfect strain transmission between the piezoelectric actuator and the $\mathrm{Fe}_{3}\mathrm{O}_{4}$ thin film, which can be described by $\Delta\epsilon_{2}^{\prime}=\chi^{\mathrm{100}}\Delta\epsilon_{2}^{\mathrm{ideal}}$ with $\chi^{\mathrm{100}}=0.27$.

%%%%%%%%%%%%%%%%%%%%%%%%%%%%%%%%%%%%%%% Fig.3 %%%%%%%%%%%%%%%%%%%%%%%%%%%%%%%%%%%%%%%%%%%%%%
%%%%%%%%%%%%%%%%%%%%%%%%%%%%%%%%%%%%%%% Fig.3 %%%%%%%%%%%%%%%%%%%%%%%%%%%%%%%%%%%%%%%%%%%%%%

The corresponding free energy contours within the film plane $F/M_{\mathrm{s}}(\Theta, \Phi=90^\circ)$ in the absence of an external magnetic field are shown in Fig.~\ref{fig:FMR-FreeE}(b). In agreement with Figs.~\ref{fig:Illustration-FreeE}(b)--(d), the contour for $V_{\mathrm{p}}=0$\,V ($\epsilon_{2}^{\prime}=0$) exhibits a fourfold symmetry with a superimposed magnetic hard axis, which is found to be along $\Theta_{\mathrm{u}}=0^{\circ}$ in the experiment. Upon the application of $V_{\mathrm{p}}\neq0$ we observe a change of the relative strength of the magnetic hard axes, as evident from the different magnitudes of the maxima of the free energy, while they retain their orientation. The easy axes---i.e., the free energy density minima---almost retain their strength, but the orientation $\Theta$ of the easy axes clearly is dependent on $V_{\mathrm{p}}$. This is the basis for the continuous and reversible rotation of $\mathbf{M}$ in the spin-mechanics scheme and confirms the simulations of Fig.~\ref{fig:Illustration-FreeE}. However, the free energy density contours in Fig.~\ref{fig:FMR-FreeE}(a) reveal a rotation of the easy axes by $\Delta\Theta=6{}^{\circ}$ for -30\,V$\leq V_{\mathrm{p}}\leq +90$\,V at room temperature. Hence, a continuous and reversible voltage control of magnetization orientation is possible, but a voltage-controlled magnetization switching is out of reach in the present hybrid, since the induced strain $\epsilon_{2}^{\prime}$ is much lower than the nominal strain of the piezoelectric actuator.  

In case of $\alpha=45^{\circ}$, the experimentally obtained and simulated FMR fields $\mu_{0}H_{\mathrm{res}}\left(\theta\right)$ are shown in Fig.~\ref{fig:FMR-FreeE}(c). In analogy to the configuration $\alpha=0^{\circ}$, the total free energy density $F$ for the present sample is composed of Eq.~(\ref{eq:F-tot_general}), with the additional, thermally induced uniaxial anisotropy $F_{\mathrm{u}}^{\mathrm{ip}}$ in the film plane along $\Theta_{\mathrm{u}}=5^{\circ}$. The solid lines in Fig.~\ref{fig:FMR-FreeE}(b) represent the numerically simulated FMR fields using the voltage-independent anisotropy fields $K_{\mathrm{u,eff}}^{001}/M_{\mathrm{s}}=75.3$\,mT, $K_{\mathrm{c}}/M_{\mathrm{s}}=-14.5$\,mT, $K_{\mathrm{u}}^{\mathrm{ip}}/M_{\mathrm{s}}=1.1$\,mT, and the strain $\epsilon_{2}^{\prime}$ as fit parameter. From these values a non-ideal strain transfer of $\chi^{\mathrm{110}}=0.09$ can be inferred. The corresponding calculated free energy density curves in the film plane $F/M_{\mathrm{s}}(\Theta, \Phi=90^\circ)$ are depicted in Fig.~\ref{fig:FMR-FreeE}(d). According to  Section~\ref{sec:Concept}, upon the application of a voltage $V_{\mathrm{p}}$, the energy minima mainly retain their orientation. More importantly, the relative strengths of the magnetic easy axes considerably change, as illustrated for the energy density minimum at $\Theta=133{}^{\circ}$, which remarkably loses depth for $V_{\mathrm{p}}=+150$\,V and approaches transforming into a maximum. Due to the low $\chi^{\mathrm{110}}$ value the strain-induced anisotropy is unfortunately not large enough to cause an abrupt magnetization switching as shown in Fig.~\ref{fig:Illustration-FreeE}(f) and (g). However, by optimizing the strain transmission efficiency, magnetization switching should be possible for $\alpha=45^{\circ}$. 

Figure~\ref{fig:FMR-FreeE} demonstrates that angular-dependent FMR measurements allow to quantitatively determine the contributions to the total free energy density $F$. However, it does not directly measure the remanent magnetization orientation. Moreover, Eq.~(\ref{eq:F-tot_general}) is applicable only to homogeneously magnetized samples. This is valid for FMR measurements, since the applied external field suffices to fully saturate the magnetization for the present hybrids. As we are particularly aiming at a magnetization switching at vanishing external magnetic field, magnetic domain formation might be important. Therefore, in the following, we utilize superconducting quantum interference device (SQUID) magnetometry measurements as a function of the in-plane magnetic field orientation $\theta$ to directly measure the remanent magnetization as a function of $V_{\mathrm{p}}$. For these angular-dependent magnetization measurements, we magnetized the hybrid along a magnetically easy axis by applying $\mu_{0}H_{\mathrm{prep}}=+1$\,T and then swept the magnetic field to $\mu_{0}H=0$\,T at a fixed strain state,~i.e.,~at a fixed voltage $V_{\mathrm{p}}$. After the preparation of the magnetization, we recorded the projection of the magnetization on the magnetic field direction as a function of the magnetic field orientation $\theta$.

%%%%%%%%%%%%%%%%%%%%%%%%%%%%%%%%%%%%%%% Fig.4 %%%%%%%%%%%%%%%%%%%%%%%%%%%%%%%%%%%%%%%%%%%%%%
\begin{figure}
\begin{centering}
\includegraphics[width=0.95\columnwidth]{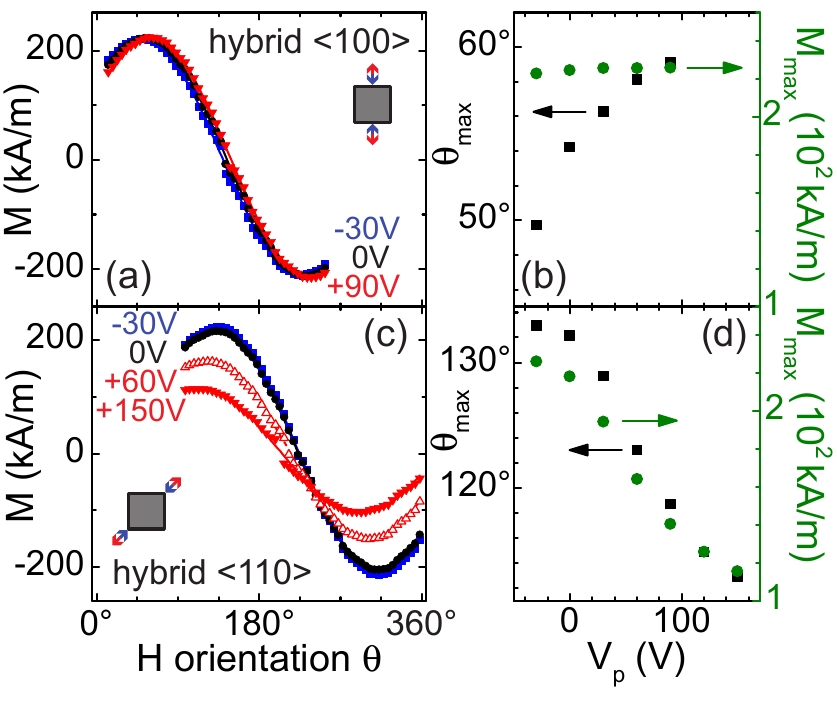}
\end{centering}
\caption{SQUID magnetometry measurements as a function of $\theta$ with $\phi=90^{\circ}$ at different voltages $V_{\mathrm{p}}$ using the hybrid~$\left\langle 100\right\rangle$ (a),~(b) and the hybrid~$\left\langle 110\right\rangle$ (c),~(d). All measurements were carried out at $\mu_{0}H=0$\,mT. The symbols represent the experimental data and the lines denote fits to cosine functions. (b),~(d) Orientation $\theta_{\mathrm{max}}$, which denotes the angle of the maximum value of $M\left(\theta\right)$ (black squares), and the corresponding magnitude $M_{\mathrm{max}}$ (green circles) as a function of $V_{\mathrm{p}}$.}
\label{fig:Mrot}
\end{figure}
%%%%%%%%%%%%%%%%%%%%%%%%%%%%%%%%%%%%%%% Fig.4 %%%%%%%%%%%%%%%%%%%%%%%%%%%%%%%%%%%%%%%%%%%%%%

In case of the hybrid~$\left\langle 100\right\rangle$, the preparation of the magnetization was carried out at $V_{\mathrm{p}}=-30$\,V with the external magnetic field oriented along $\theta_{\mathrm{prep}}=50^{\circ}$, which corresponds to a minimum of the free energy density $F$ [cf.~Fig.~\ref{fig:FMR-FreeE}(b)]. The results obtained by carrying out angular-dependent magnetometry measurements are shown in Fig.~\ref{fig:Mrot}(a). Since in the absence of an external magnetic field the magnetization preferably aligns along a magnetic easy axis, maxima in the $M\left(\theta\right)$ curves correspond to minima in the $F\left(\Theta\right)$ contours. The respective maxima of the $M\left(\theta\right)$ curves are evaluated in Fig.~\ref{fig:Mrot}(b), regarding their orientation $\theta_{\mathrm{max}}$ (black squares) and magnitude $M_{\mathrm{max}}$ (green circles) as a function of $V_{\mathrm{p}}$. $M_{\mathrm{max}}$ changes by only 1\% in the voltage range -30\,V$\leq V_{\mathrm{p}}\leq +90$\,V and thus is almost independent of $V_{\mathrm{p}}$. This demonstrates that domain formation plays only a negligible role in case of $\alpha=0^{\circ}$. However, $\theta_{\mathrm{max}}$ changes by about $9^{\circ}$. This proves that the macroscopic, homogeneous remanent magnetization $\mathbf{M}$ rotates by about $9^{\circ}$ in the film plane for -30\,V$\leq V_{\mathrm{p}}\leq +90$\,V, which confirms the results obtained by FMR measurements [cf.~Fig.~\ref{fig:FMR-FreeE}(b)].

We now turn to the hybrid~$\left\langle 110\right\rangle$. In a first set of experiments, the preparation field was applied along $\theta_{\mathrm{prep}}=133^{\circ}$ [Fig.~\ref{fig:Mrot}(c)]. As the free energy density at this orientation continuously evolves from a deep minimum towards a shallow one with increasing $\epsilon_{2}^{\prime}$,~i.e.,~$V_{\mathrm{p}}$ [cf.~Fig.~\ref{fig:FMR-FreeE}(d)], this minimum will be referred to as local minimum in the following. The angle-dependent SQUID measurements reveal a qualitatively different behavior compared to the measurements on the hybrid~$\left\langle 100\right\rangle$. In case of the hybrid~$\left\langle 110\right\rangle$, both the magnitude $M_{\mathrm{max}}$ as well as the orientation $\theta_{\mathrm{max}}$ of the maximum significantly change as a function of $V_{\mathrm{p}}$ [cf.~Fig.~\ref{fig:Mrot}(d)]. Upon increasing $V_{\mathrm{p}}$ from -30\,V to +150\,V, $M_{\mathrm{max}}$ decreases by 49\% of its initial value, while the orientation of $M_{\mathrm{max}}$ rotates by $20{}^{\circ}$ towards the free energy density minimum at $47{}^{\circ}$ [cf.~Fig.~\ref{fig:Illustration-FreeE}(f)]. The reduction of $M_{\mathrm{max}}$ elucidates magnetic domain formation with increasing $V_{\mathrm{p}}$. After the magnetic preparation at $V_{\mathrm{p}}=-30$\,V, the $\mathrm{Fe}_{3}\mathrm{O}_{4}$ thin film exhibits a single-domain state with the homogeneous magnetization oriented along $\Theta=133^{\circ}$. As the applied voltage $V_{\mathrm{p}}$ increases, the local energy density minimum at $\Theta=133^{\circ}$ looses depth and thus magnetically hardens, while the global minimum of the energy density at $\Theta=47^{\circ}$ magnetically softens, favoring domain formation [cf.~Fig.~\ref{fig:FMR-FreeE}(c)]. Thus, the angular-dependent magnetization measurements shown in Figs.~\ref{fig:Mrot}(c) and (d) are not consistent with the single-domain free energy density approach used to calculated the energy density contours in Fig.~\ref{fig:FMR-FreeE}(d).        

In a second set of experiments, we repeated the SQUID measurements with the magnetic preparation field $\mathbf{H}_{\mathrm{prep}}$ applied along $\theta_{\mathrm{prep}}=43^{\circ}$, close to the global minimum of the free energy density at $\Theta=47^{\circ}$. The experimental data coincide in good approximation for different applied voltages $V_{\mathrm{p}}$  [not shown here]. The magnetization $\mathbf{M}$ retains its orientation at $\Theta=47^{\circ}$ independent of $V_{\mathrm{p}}$, while the magnitude of the magnetization at this orientation $M_{\mathrm{max}}$ changes by only 5\% within the full voltage range. Considering the free energy density contours [cf.~Fig.~\ref{fig:FMR-FreeE}(d)], the energy barrier for domain formation is much larger in this case, such that the $\mathrm{Fe}_{3}\mathrm{O}_{4}$ thin film remains in a magnetically single domain state and can be described by Eq.~(\ref{eq:F-tot_general}).

\section{Impact of different strain orientations}
\label{sec:strain orientation}

The experimental results discussed above show that an alignment of the strain axes along crystallographic axes might favor magnetic domain formation. Therefore, we now discuss configurations with an angle $\alpha$ in between $0^{\circ}$ and $45^{\circ}$.
%%%%%%%%%%%%%%%%%%%%%%%%%%%%%%%%%%%%%%% Fig.5 %%%%%%%%%%%%%%%%%%%%%%%%%%%%%%%%%%%%%%%%%%%%%%
\begin{figure}[tbh]
\begin{centering}
\includegraphics[width=0.95\columnwidth]{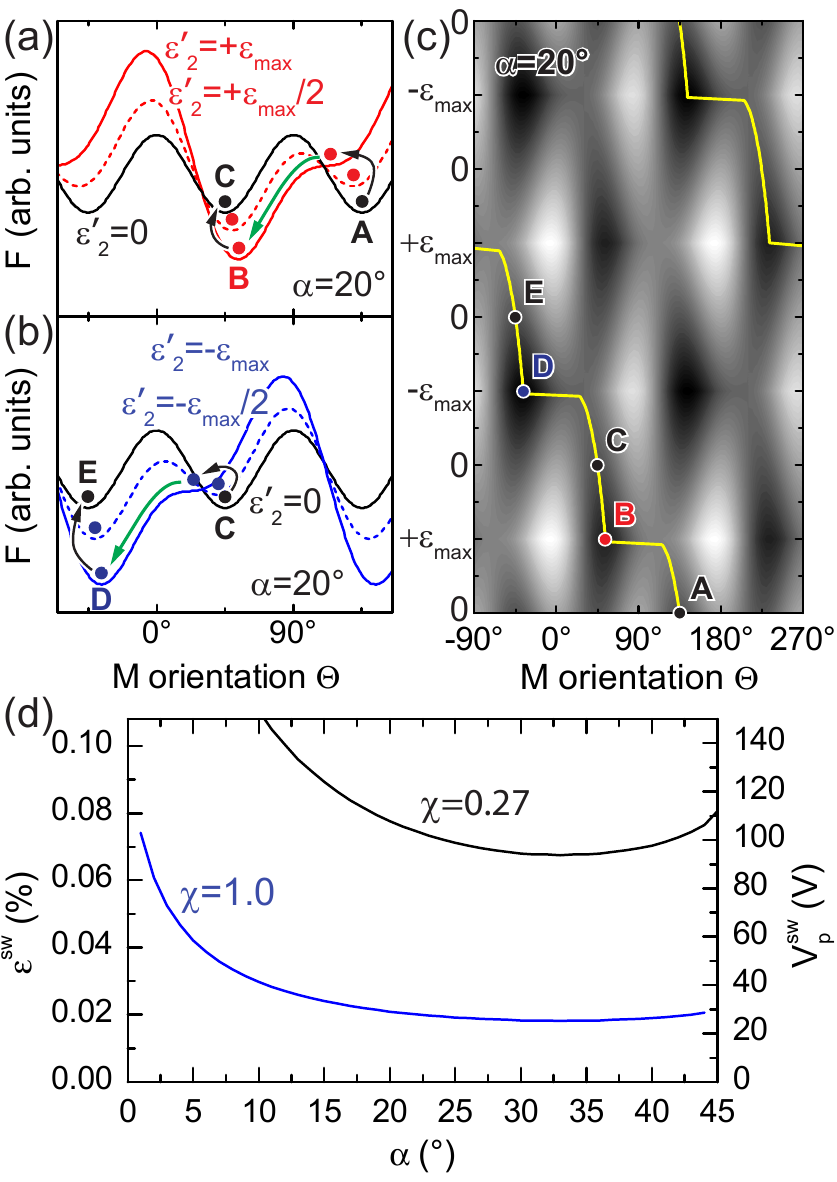}
\end{centering}
\caption{Approach to a nonvolatile, all-voltage controlled magnetization switching. (a)--(c) Calculated free energy density contours $F\left(\Theta,\epsilon_{2}^{\prime}\right)$ for $\alpha=20{}^{\circ}$. The full
yellow line in (c) traces the minimum of $F$. Discontinuous switching processes of the magnetization by $90^{\circ}$ from $\mathrm{A} \rightarrow \mathrm{C} \rightarrow \mathrm{E}$ while applying $\epsilon^{\mathrm{sw}}$ with alternating sign are visible. (d) Calculated critical strain $\epsilon^{\mathrm{sw}}\left(\alpha\right)$ values and corresponding switching voltages $V_{\mathrm{p}}^{\mathrm{sw}}\left(\alpha\right)$ at which a discontinuous switching of the magnetization, which is illustrated by the green arrows in (a) and (b), occurs. The blue curve depicts perfect strain transmission ($\chi=1.0$) and the black curve represents the experimentally realized strain in the hybrid~$\left\langle 100\right\rangle $ ($\chi=0.27$).} 
\label{fig:switching}
\end{figure}
%%%%%%%%%%%%%%%%%%%%%%%%%%%%%%%%%%%%%%% Fig.5 %%%%%%%%%%%%%%%%%%%%%%%%%%%%%%%%%%%%%%%%%%%%%%
The corresponding concept is exemplarily illustrated for $\alpha=20^{\circ}$ in Figs.~\ref{fig:switching}(a), (b), and (c). 

Starting at $\epsilon_{2}^{\prime}=0$, we assume an initial magnetization orientation along $\Theta=135^{\circ}$ [point A in Fig.~\ref{fig:switching}(a)]. Upon increasing $\epsilon_{2}^{\prime}>0$ in the thin film, the easy axis continuously rotates, until it switches discontinuously to point B at a certain critical strain $\epsilon_{2}^{\prime,\mathrm{crit}}=\epsilon^{\mathrm{sw}}\left(\alpha\right)$ [green arrow in Fig.~\ref{fig:switching}(a)]. When the strain $\epsilon_{2}^{\prime}$ is reduced back to $0$, the easy axis rotates to point C at $\Theta=45^{\circ}$. Upon subsequently increasing the strain $\epsilon_{2}^{\prime}<0$ with opposite sign [Fig.~\ref{fig:switching}(b)], the situation appears qualitatively different from the situation illustrated in Fig.~\ref{fig:Illustration-FreeE}(g), as we do not observe a back switching to the initial orientation (point A), but a further switching process along the original direction of rotation via point D to point E at $\Theta=-45{}^{\circ}$ [Fig.~\ref{fig:switching}(b)]. Hence, iteratively applying $\epsilon^{\mathrm{sw}}$ with alternating sign provides a concept to discontinuously rotate the equilibrium magnetization orientation by $90^{\circ}$ via nonvolatile switching processes.\cite{Iwasaki:JMMM:240:2002} Such magnetization switching processes are ``quasi-reversible'', since four consecutive switching processes (in a ferromagnet with cubic symmetry) evidently restore the initial magnetization orientation state [Fig.~\ref{fig:switching}(c)]. Hence, this constitutes a very elegant voltage-control scheme of magnetization orientation.
    
Assuming a perfect strain transmission between the piezoelectric actuator and the ferromagnetic thin film ($\chi=1$), $\epsilon^{\mathrm{sw}}$ can be derived using the free energy density $F$ given in Eq.~(\ref{eq:F-tot_general}) with a cubic anisotropy field $K_{\mathrm{c}}/M_{\mathrm{s}}=-14.7$\,mT, which is the averaged value of the cubic anisotropy measured in hybrid~$\left\langle 100\right\rangle$ and hybrid~$\left\langle 110\right\rangle$, as well as the elastic and magnetoelastic constants of $\mathrm{Fe}_{3}\mathrm{O}_{4}$. Figure~\ref{fig:switching}(d) shows that $\epsilon^{\mathrm{sw}}$ required to induce a magnetization switching process significantly decreases with increasing angle $\alpha$, exhibits a minimum at $\alpha=33^{\circ}$, and finally slightly increases with $\alpha$ approaching $45^{\circ}$. Overall, $\epsilon^{\mathrm{sw}}$ has comparatively moderate values lower than $10^{-3}$, which are experimentally achievable using the concept described in Fig.~\ref{fig:piezo-coordsyst}(a). These values correspond to switching voltages $V_{\mathrm{p}}^{\mathrm{sw}}\left(\alpha\right)<150$\,V in our hybrid concept, which are experimentally accessible [cf.~Fig.~\ref{fig:switching}(d)]. 

To furthermore lower the switching strain $\epsilon^{\mathrm{sw}}$, the properties of the ferromagnetic film itself must be fine-tuned, as $\epsilon^{\mathrm{sw}}$ linearly depends on the cubic anisotropy constant $K_{\mathrm{c}}$ and inversely depends on the magnetostriction constants $\lambda_{100}$ and $\lambda_{111}$. Most promising candidates regarding the realization of a magnetization switching therefore evidently are materials with a small cubic anisotropy and high magnetostriction constants.

\section{Conclusion}
\label{sec:Conclusion}

In summary, we have investigated concepts for a voltage-controlled, nonvolatile $90{}^{\circ}$ switching of the remanent magnetization in $\mathrm{Fe}_{3}\mathrm{O}_{4}$ thin film/piezoelectric actuator hybrids at room temperature. The possibility to induce strain along different directions in the film plane with respect to the crystallographic axes depending on the cementing procedure, allows to investigate the switching behavior and particularly to take advantage of the magnetostriction constants $\lambda$ along different crystalline orientations. We have discussed the qualitatively different switching behavior for two different configurations, namely strain exerted along the in-plane crystalline $\mathrm{Fe}_{3}\mathrm{O}_{4}$ $\left\langle 100\right\rangle $ and along the in-plane $\left\langle 110\right\rangle $ directions. The free energy density of the ferromagnetic thin films was determined by FMR spectroscopy, which allows to infer the equilibrium magnetization orientation in a Stoner-Wohlfarth model. The results show a rotation of the easy axes by a few degrees and a significant modification of the relative strength of the easy axes, respectively. However, in combination with SQUID magnetometry measurements we find that the angle of magnetization reorientation is not large enough to induce a magnetization switching in the former, and magnetic domain formation impedes a coherent magnetization switching in the latter approach. This shows that inefficient strain transfer and magnetic domain formation are major obstacles towards a non-volatile strain-controlled magnetization switching. Using the experimental free energy and strain transfer parameters, we find in simulations that skillful alignment of the strain within the films should reduce the strain values required to switch the magnetization and impede domain formation. More explicitly, our experiments suggest that an all-voltage-controlled, nonvolatile magnetization switching at room temperature and zero magnetic field should be possible.

Financial support via DFG Project No. GO 944/3-1 and the German Excellence
Initiative via the ``Nanosystems Initiative Munich (NIM)'' are gratefully
acknowledged.

\end{document}